\documentclass[pdflatex,sn-mathphys-num]{sn-jnl}

\usepackage{graphicx}%
\usepackage{multirow}%
\usepackage{amsmath,amssymb,amsfonts}%
\usepackage{amsthm}%
\usepackage{mathrsfs}%
\usepackage[title]{appendix}%
\usepackage{xcolor}%
\usepackage{textcomp}%
\usepackage{manyfoot}%
\usepackage{booktabs}%
\usepackage{algorithm}%
\usepackage{algorithmicx}%
\usepackage{algpseudocode}%
\usepackage{listings}%
\usepackage{bm}
\usepackage{tikz}

\begin{document}
	
	\title{Effects of high-order Van Hove singularities on exciton and trion energy dispersions}

	\author*{\fnm{Lewis J.} \sur{Burke}}\email{L.J.Burke@icloud.com}
	
	\author*{\fnm{Mark T.} \sur{Greenaway}}\email{M.T.Greenaway@lboro.ac.uk}
	
	\author*{\fnm{Joseph J.} \sur{Betouras}}\email{J.Betouras@lboro.ac.uk}
	
	\affil{Department of Physics and Centre for the Science of Materials, Loughborough University, LE11 3TU, United Kingdom}

	\abstract{We investigate the effects of Van Hove singularities in the electronic band structure of two-dimensional semiconductors on the energy dispersion of excitons and positive trions. In particular, we study valence band energy dispersions which possess (i) a typical logarithmic Van Hove singularity, (ii) a line high-order Van Hove singularity (HOVHS) from a Mexican-hat dispersion or (iii) a point HOVHS such as a monkey saddle.  We find that the density of states (DOS) of excitons and trions containing such singularities is dramatically enhanced and shows, in general, how the HOVHS in the valence band can strongly affect and be mirrored in the DOS of excitons and trions. This leads to new states that govern the optical properties of the system. In addition, we study a set of materials, InSe, GaSe and $\alpha$-SnAs, from a class of materials in which the topmost valence band has an inverted Mexican-hat shape. The most favourable exciton occurs when the singularity is at the $\Gamma$-point, as in the example of monolayer $\alpha$-SnAs, which hosts a HOVHS.  Our work thus provides a pathway to engineer specific bound states in two-dimensional materials that host such singularities, thereby opening new avenues for potential applications.}
	
	\maketitle

	\section{Introduction} Excitons and trions are electrostatically bound quasiparticles generated by photon absorption \cite{frenkel1931transformation}. They play a crucial role in the electronic and optical properties of a quantum material. An exciton is neutral in charge and consists of a bound electron-hole pair, whereas a trion (also known as a charged exciton) forms in the presence of low electron or hole doping, where an initial exciton binds to the doping charge, creating a three-particle complex \cite{lampert1958mobile, esser2001theory}.  At higher doping levels, many other many-body quasiparticles, such as Fermi polarons, should also be considered \cite{muir2022interactions}.
	These quasiparticles have been intensively studied in a variety of two-dimensional (2D) materials from transition metal dichalcogenides (TMDs) \cite{mueller2018exciton,Rana2016,Vaclavkova_2018,Tempelaar_2019,shradha20262d}, to other hexagonal chalcogenides such as InSe \cite{Falko2020,burke2025brightening}. They can also form in bilayer (and multilayer) systems, in which interlayer excitons as well as moiré excitons and trions form \cite{brotons2021moire,schmitt2022formation}.
	
	The band structure of the material is critical in the formation of excitons and trions. In particular, the existence of flat bands or Van Hove singularities (VHSs) in the Brillouin zone (BZ) influences and, to a good extent, governs the physical behaviour of the system. Earlier work has demonstrated that a VHS enhances electron-hole (excitonic) correlation in optical spectra \cite{PhysRevLett.103.186802,PhysRevLett.106.046401}. Recently, the role of VHSs on other quasiparticles, such as polaritons, has been studied \cite{gianardi2025formation}.  In this work, we investigate how the presence of either a conventional VHS or a high-order Van Hove singularity (HOVHS) in the valence band directly influences the energy dispersions of excitons and trions that form in intrinsic or doped 2D semiconductors. 
	
	VHSs appear in the electronic band structure of a system when the gradient of the energy dispersion, $\nabla_k E(\bm{k})$, vanishes at the Fermi surface \cite{vanHove}. In 2D systems, a conventional VHS leads to a logarithmic singularity in the DOS when $\nabla_k E(\bm{k}) =0$ but the Hessian determinant is negative. HOVHSs occur when the determinant of the Hessian of the energy dispersion also vanishes \cite{ChandrasekaranVHS}. The immediate consequence is that the density of states (DOS) diverges with a power-law, characteristic of the type of the HOVHS as the critical energy is approached.
	
	HOVHSs play an important role in phase formation, thermodynamics and transport properties of quantum materials \cite{classen2024high, chandrasekaran2023engineering, chandrasekaran2023practical, Efremov_Betouras2019, Sherkunov_Betouras2018}. However, their effects on the properties of excitons and trions are not well understood. Previously, \cite{burke2025brightening}, we showed that the underlying band structure strongly influences the binding energy of the trions. Our calculations on monolayer InSe revealed that the negative trion, which comprises two electrons in the parabolic conduction band, is bound. However, the positive trion, which comprises holes at the band edge of a Mexican-hat dispersion that hosts a VHS, is unbound. The reason for this is, in part, the enhanced hole-hole interaction due to the VHS of the spin-degenerate valence bands, as well as the increase in kinetic energy. In this work, we investigate the effects of HOVHS in the valence band on excitons and positive trions, while using a parabolic conduction band with a fixed electron effective mass. We consider the effect of different types of saddle points and show how the VHS type is mirrored within the dispersions of the exciton and positive trion states. Then, we show how the singularities formed in an inverted Mexican hat valence band energy dispersion, affect the properties of positive excitons and show how to tune the properties of the quasi particles by modifying the valence band.  We consider three materials, InSe, GaSe and SnAs as exemplary materials of the type.   We follow the computational details as described in the methods section.
	
	\section{Results}
	\subsection{HOVHSs in exciton dispersion} 
	
\begin{figure}[t!]
		\centering
		\includegraphics[width=0.8\linewidth]{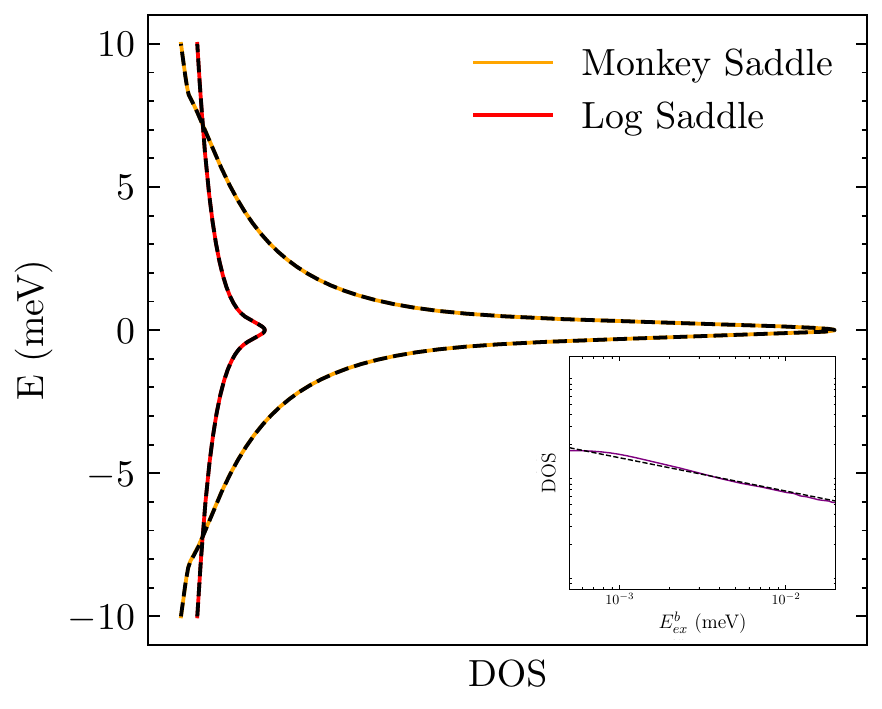}
		\caption{The DOS of the calculated ground state exciton dispersion (dashed black lines) overlaid with the DOS of the valence bands with monkey-saddle (yellow) and logarithmic saddle dispersions (red) of each system. Inset: log-log plot of the calculated DOS in the vicinity of the singularity, black dashed line is a line with gradient -1/3 highlighting the power law divergence of the monkey-saddle exciton DOS.}
		\label{fig:val_dos_VHSs}
	\end{figure}

	To understand the effect of different types of valence band singularities on the exciton properties, we first consider the valence-band dispersion relation \[ E_v(\bm{k}) = \beta(k_x^2-k_y^2)\]
	which has a saddle point at $k_x=k_y=0$, and a conventional VHS with a logarithmic divergence of the DOS, $g(E)\propto\ln(1/|E|)$. We compare this to a monkey-saddle dispersion,
	\[ E_v(\bm{k})=\beta(k_x^3-3k_xk_y^2),\]
	with a HOVHS at the $\Gamma$-point, and a power law divergence of the DOS, $g=|E|^{-1/3}$ \cite{ChandrasekaranVHS, chandrasekaran2020catastrophe}. In both cases, we take $\beta=1$, with the units eV\AA$^2$ and eV\AA$^3$ respectively.
	In Fig. \ref{fig:val_dos_VHSs}, we show the DOS of each valence band (solid curves), overlaid with the exciton DOS (dashed curves), calculated numerically using the method described in the Methods section, which use the same range of valence-band and exciton momenta. We find that the exciton and the valence band DOS peaks have the same form, so that the divergent behaviour of the valence band is mirrored in the exciton dispersion (in the inset of Fig.\ref{fig:val_dos_VHSs}, we highlight the exact power law divergence -1/3 by the dashed line which shows good comparison to the calculated DOS). 
	
	This result can be understood when we look at the non-interacting part of the exciton eigenvalue expression, see equation \ref{eq-BSE2}. Here, we write the valence band term as $E_v(\bm{k}-\bm{Q})$, where $\bm{Q}$, is the quasiparticle momenta. The excitonic dispersion $E_{ex}(\bm{Q})$ follows the non-interacting expansion of $\sum_{\bm{k}}(E_c(\bm{k})-E_v(\bm{k}-\bm{Q}))|\phi(\bm{k})|^2$, in which $\phi(\bm{k})$, is the excitonic wave function. 
	
	Expanding, and noting that since $\phi$ inherits the symmetry of the monkey saddle, $\sum_k k_x^2|\phi({\bm k})|^2$=$\sum_k k_y^2|\phi({\bm k})|^2$, $\sum_k k_xk_y|\phi({\bm k})|^2=0$ and that the odd powers of $k$ sum to zero; we find that the $\bm{Q}$ dependence on the exciton energy can be written simply as $E_{ex}(\bm{Q})=E_0-Q_x^3+3Q_xQ_y^2$. Therefore, the exciton dispersion dependence on $\bm{Q}$ has the same polynomial dependence as the valence band, and a HOVHS singularity remains at $\bm{Q}=0$ with the corresponding enhancement of the density of states.  In contrast, for the logarithmic VHS dispersion $E_v(\bm{k}-\bm{Q})= (k_x-Q_x)^2-(k_y-Q_y)^2$ so that for the term $\sum_{\bm{k}} E_v(k-Q)|\phi(k)|^2$, the odd powers of $k$ sum to zero, leading to $\sum_k (k_x^2-k_y^2+Q_x^2-Q_y^2)|\phi(k)|^2$, resulting in a  logarithmic VHS in the excitonic band.

	This analysis suggests that singularities in the valence band of a system will be mirrored within the dispersion of the exciton.  However, we find that whether the nature of the singularity is preserved in the quasi-particle dispersion is not straightforward and depends on the form of the singularity. For example, consider a Mexican-hat dispersion, which can be described simply by 
	\begin{equation} 
		E_v(\bm{k})=A\bm{k}^2+B\bm{k}^4,
		\label{eq:mexhatdisp}
	\end{equation}
	where $A>0$ and $B<0$. A material with this class of dispersion, such as the chalcogenides \cite{zolyomiBandStructureOptical2013a,Falko2016}  is an indirect semiconductor with the valence band maximum (VBM) occurring at finite $k_{max}=\sqrt{A/2|B|}$. At the VBM, there is a Van Hove line singularity, which is one-dimensional in nature and has a power-law divergence (like HOVHSs) of $|E|^{-1/2}$. Here, the expansion of $E_v(\bm{k}-\bm{Q})$ in the BSE for the exciton gives
	\[
	E_{ex}({\bm Q})=\langle E_v(\bm{k})\rangle +(A+4B\langle \bm{k}^2\rangle)\bm{Q}^2+B\bm{Q}^4 \]
	where $\langle \bm{k}^2\rangle=\sum_{\bm{k}} {\bm k}^2|\phi({\bm k})|^2$ and $\langle E(\bm{k})\rangle=\sum_{\bm{k}} E_v(\bm{k})|\phi({\bm k})|^2$ are constant terms.  Therefore, the exciton has a Mexican hat dispersion (with a shifted radius in ${\bm k}$ and a $|E|^{-1/2}$ singularity) if the coefficients of ${\bm Q}^2$ and ${\bm Q}^4$ have opposite signs, i.e. if $A+4B\langle \bm{k}^2\rangle>0$. We can rewrite the condition to find a Mexican hat in the exciton dispersion in terms of $k_{max}$ so that  
	\begin{equation}
		\langle {\bm k}^2\rangle < \frac{1}{2}k_{max}^2 .
	\end{equation}
	We note that $\langle \bm{k}^2\rangle$ is effectively the mean-square momentum spread of the exciton wavefunction.  If we consider the 1s hydrogenic wave function, we find that \[
	\langle \bm{k}^2\rangle=\frac{k_c^2}{2+k_c^2a^2}
	\]
	where $a$ is the exciton radius and the Brillouin zone size is defined by $k_c=d k_{max}$, where $d\geq1$ measures the distance between the VBM and the Brillouin zone edge. The condition for a Mexican hat dispersion is then 
	\begin{equation}\begin{split}
			\sqrt{2\left(1-\frac{1}{d^2}\right)}\frac{1}{k_{max}}<a.
			\label{eq:condition}
		\end{split}
	\end{equation}
	This shows that for a range of $k_{max}$ and $k_c$ where this condition holds, there is a bound exciton with a Mexican hat dispersion.  This suggests that the line singularity only exists if the exciton is sufficiently localised in $k$-space (extended in real space) relative to the Mexican hat brim radius.  Conversely, an exciton that is tightly bound in real space (to give a large $\langle {\bm k}^2\rangle$) has a typical parabolic form and the $|E|^{-1/2}$ divergence is lost.  In particular, removing the parabolic term ($A=0$, so $k_{max}=0$) gives a purely quartic valence band, yet the exciton still has a parabolic term $4B\langle {\bm k}^2\rangle{\bm Q}^2$. The condition in Eq. \ref{eq:condition} then cannot be met, so the exciton has a parabolic component breaking the mirroring of the valence band.
	
	\subsection{Trion formation in materials with VHSs} 
	
\begin{figure}
		\centering
		\includegraphics[width=0.75\linewidth]{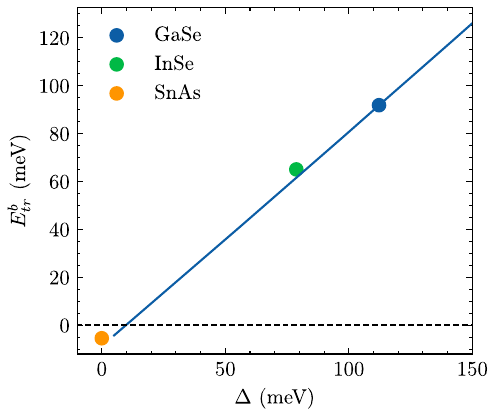}
		\caption{Trion binding energies, $E^b_{tr}$, for three materials, GaSe, SnAs, and InSe (as indicated by the legend), as a function of their Mexican hat depths $\Delta$. The blue curve shows the change in trion binding energy with Mexican-hat depth, $\Delta$ of the GaSe system with fixed $k_{max}$. The horizontal line indicates when the trion goes from bound (positive energies) to unbound (negative energies).}
		\label{fig:materials}
	\end{figure}

	We now consider trion formation for systems with Mexican hat valence bands and consider the role of the Van Hove singularities in their formation.  As an example, we consider the group III–VI hexagonal monochalcogenides, $\text{In}_2\text{Se}_2$ \cite{Falko2016,muddDirecttoindirectBandGap2016} and $\text{Ga}_2\text{Se}_2$ \cite{ben2018valence}, which are materials both known to host inverted Mexican hat valence bands.  Comparison of the two materials reveals that GaSe has a smaller electron effective mass $(m_{GaSe}=0.16, m_{InSe}=0.188)$ and an increased Mexican-hat depth ($\Delta_{GaSe}=112\text{ meV}, \Delta_{InSe}=79\text{ meV}$). We use these parameters to calculate the binding energy of the positive trion that might form in the valence band of GaSe and InSe, blue and green dots in Fig \ref{fig:materials}.  We find that the positive trion is unbound, meaning that the overall trion binding energy in these materials is lower than that of the exciton.  This is in contrast to the negatively charged trion, which is bound in InSe, as we explored in \cite{burke2025brightening}. 
	
	We find that the stronger binding of the exciton in InSe compared to that in GaSe is primarily due to its relatively shallower Mexican hat.  To show this dependence, in Fig. \ref{fig:materials} we plot the trion binding energy as a function of the Mexican hat depth, $\Delta$.  We find that as the Mexican hat depth increases, the trion's binding energy (defined relative to the exciton binding energy) decreases (becomes more positive), which means that the positive trion becomes less energetically favourable compared to the exciton. We show this by keeping the width of the Mexican hat fixed to that of GaSe, $k_{max}=0.232$\AA$^{-1}$, and varying $\Delta$.
	This trend suggests that the most favourable situation for the positive trion would occur when $A=0$. We note that the parabolic component of a valence band at the $\Gamma$ point is generally not forbidden by symmetry: time-reversal and spatial inversion symmetry constrain the dispersion to be even in ${\bm k}$, and therefore allow quadratic terms. Nevertheless, bands with $A\approx 0$ can arise in materials due to hybridisation with nearby bands.
	
\begin{figure*}[t!]
		\centering
		\includegraphics[width=\linewidth]{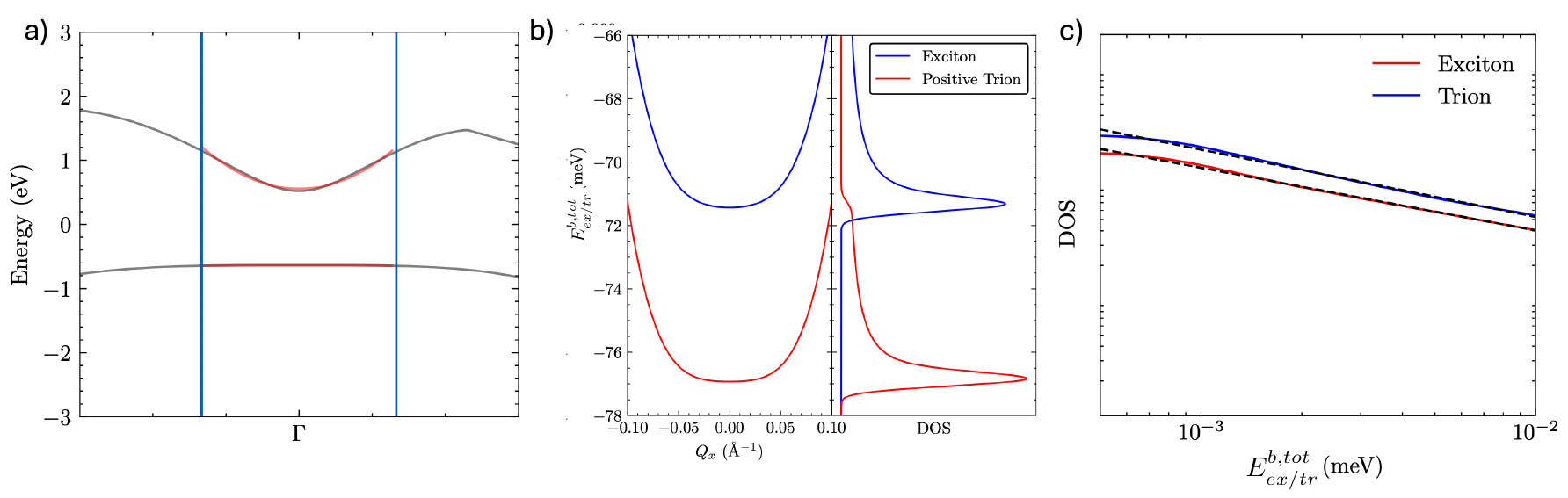}
		\caption{a) Band structure of $\alpha$-SnAs calculated using PBE-DFT close to the $\Gamma$-point (black curve). The fitted polynomials are indicated in red (see explaination in text). b) Calculations of the exciton (blue) and positive-trion (red) dispersions and corresponding DOS for doped monolayer SnAs. c)
			Log-log plot of the DOS in panel (b), highlighting the divergence, the dashed curves show the expected analytical result of a gradient $-1/2$.}
		\label{fig:SnAs}
	\end{figure*}

	There is an adjacent set of group IV–V monolayer materials, which have an inverted Mexican-hat dispersion in the valence band. Most interestingly for this study, is the material $\alpha$-SnAs (in the configuration $\text{Sn}_2\text{As}_2$),  which has been shown to have a valence band dispersion which is almost purely quartic in its power dependence on the magnitude of the wavevector ${\bm  k}$ \cite{ozdamar2018structural}, with a HOVHS at the $\Gamma$ point.
	In Fig. \ref{fig:SnAs}a, we show the results of the density functional theory (DFT) calculations for the valence and conduction bands (for the band structure, we use the PBE energy functional and the energy cutoffs for the wave function and charge density of 90 Ry and 720 Ry, respectively, on a k-point grid of 21×21×1)
	\cite{QE-2009,QE-2017,doi:10.1063/5.0005082}.  The red curve in Fig.\ref{fig:SnAs}a, shows the best fit of a quartic valence band ($E(\bm{k})=B\bm{k}^4$) and a parabolic conduction band ($E(\bm{k})=\hbar^2\bm{k}^2/2m$),
	where we obtain $B\approx -50\text{eV\AA}^2$ (in agreement with \cite{ozdamar2018structural}) and an electron effective mass $m_e=0.072$ (a value close to other similar materials, e.g. in monolayer GaAs, the electron effective mass at the $\Gamma$-point is $\approx$ 0.07 \cite{hasani2022effects}). 
	
	The quartic polynomial in the valence band is of particular interest as it corresponds to a flattened band which hosts a HOVHS with a power-law divergence in the DOS. For an energy dispersion $E(\bm{k})=\bm{k}^n$, the corresponding divergence of the DOS approaching the critical point is \cite{usui2017enhanced, chandrasekaran2020catastrophe}:
	\begin{equation}
		g(E)\propto|E|^{\frac{d-n}{n}}
	\end{equation}
	where $d=2$ corresponds to the dimensionality of the system. Thus, 
	for $n=4$ we obtain a strong divergence of $|E|^{-1/2}$ (so that a dispersion $B\bm{k}^4$ gives rise to a DOS scaling of $(1/\sqrt{B})|E|^{-1/2}$).
	The ground-state exciton and positive trion binding energies for $\alpha$-SnAs are calculated under the assumptions described in the Methods section and the same used for the InSe and GaSe calculations, see Fig. \ref{fig:SnAs}(b). The curves reveal that both the exciton (blue) and trion (red) dispersions are very flat, but as previously discussed, they are not purely quartic as their valence band counterparts, but rather have a quadratic term in $\bm{Q}$. However, a polynomial fit in $\bm{Q}$ reveals only a small quadratic contribution to both dispersions, which are of the order, respectively, 350 and 700 times smaller than the coefficient of the quartic term.  This means that both dispersions are, to a very good approximation, quartic and that the trion dispersion is slightly flatter than the exciton.  
	
	The calculations also reveal that the ground state of the positive trion is a bound state, with a larger amplitude binding energy relative to the exciton due to its flatter valence band. The higher DOS peak corresponding to the trion is a result of the trion's energy dispersion being slightly flatter than that of the exciton. This is because the curvature of the band is inversely dependent on the effective mass of the quasiparticle; the trion, having a larger effective mass (expected due to the existence of a doping charge, $2m_h+m_e>m_h+m_e$), directly leads to a flatter dispersion \cite{meneghini2026arpes}.  In Fig. \ref{fig:SnAs}c, we show that the divergence of the DOS of both the exciton and trion for $\alpha$-SnAs closely approaches a power law divergence of $-1/2$, expected of a quartic dispersion and mirroring that of the valence band, with the associated relative enhancement of the DOS. 
	As a result, we see a strong effect of the HOVHS in the valence band in the DOS of the exciton and similarly for the trion.
	
\begin{figure}[ht]
		\centering
		\includegraphics[width=0.75\linewidth]{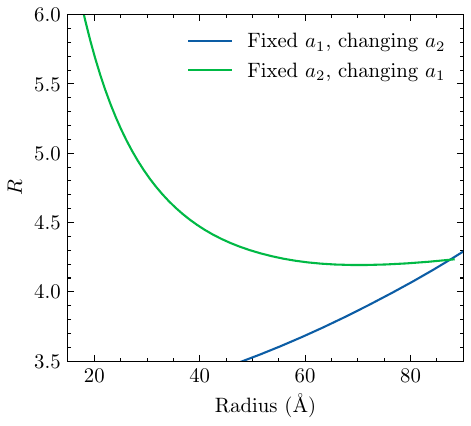}
		\caption{Plot of the ratio, $R$, between the hole-hole interaction and electron-hole interactions varying $a_1$ and $a_2$ with fixed $a_2$ and $a_1$ respectively The fixed values are the numerically obtained values from InSe, $a_1=15$ \AA, $a_2=50$ \text{\AA}  \cite{burke2025brightening}.}
		\label{fig:int}
	\end{figure}

	This result shows that band flattening has a pronounced effect on the exciton and positive trion states.  Moving from a line HOVHS to a point HOVHS leads to an energetically more favourable trion state. Therefore, indicating that the width of the Mexican hat (movement of the singularity) also has a pronounced effect on the trion binding energy, along with the depth. The reduction of binding energy as an increase in $\Delta$ can be thought of as increased kinetic energy over a purely flat band. In addition, the repulsive interaction between the alike charges here can also be attributed to the lowering in binding energy. This influence of the Coulomb interaction for the trion can be seen by investigating the changes in hole-hole and electron-hole interactions as the radius values vary (see Fig. \ref{fig:int}). We define two radii $a_1$ and $a_2$, where $a_1$ corresponds to the electron-hole pair present for the exciton, whilst $a_2$ corresponds to the outer hole generated by hole doping. These are present in the trion wave function $\phi_{tr}(\bm{k}_1,\bm{k}_2)$, where, following Eq. \ref{Eq-trionwavefn}, each $\phi(\bm{k_i})$ corresponds to the exciton 1s wave function in  Eq. \ref{eq-1s_anisotropic}, and we define from Eq. \ref{Eq-trionwavefn} $a_1=1/\alpha$ and $a_2=1/\gamma$.  
	
	To understand these competing electrostatic forces, we compare the ratio of electron-hole ($e-h$) to the hole-hole ($h-h$) interactions, as defined by:\begin{equation}
		\begin{split}
			\langle V_{h-h}\rangle&=\sum_{\bm{k}_1,\bm{k}_2}\phi^*_{tr}(\bm{k}_1,\bm{k}_2)\sum_{\bm{q}} V(\bm{q})\phi_{tr}(\bm{k}_1+\bm{q},\bm{k}_2-\bm{q})\\ \langle V_{e-h}\rangle&=\sum_{\bm{k}_1,\bm{k}_2}\phi^*_{tr}(\bm{k}_1,\bm{k}_2)\left[\sum_{\bm{q}} V(\bm{q})\phi_{tr}(\bm{k}_1-\bm{q},\bm{k}_2)+\sum_{\bm{q}} V(\bm{q})\phi_{tr}(\bm{k}_1,\bm{k}_2-\bm{q})\right]
		\end{split}
	\end{equation} 
	where the ratio is $R=\langle V_{e-h}\rangle/\langle V_{h-h}\rangle$.  
	We consider how $R$ varies when i) we fix $a_1$ and change $a_2$ (blue curve in Fig. \ref{fig:int}) and ii) we fix $a_2$ and change $a_1$ (orange curve in Fig. \ref{fig:int}). We find that as $a_2$ increases, $R$ increases. This indicates that the relative strength of the electron-hole attraction grows as the outer hole spreads out, suppressing the repulsive hole-hole term, leading to a more favourable trion state. Conversely, as $a_1$ increases for fixed $a_2$, $R$ decreases; because the holes become closer in proximity, causing the repulsive hole-hole interaction to become more substantial and the overall trion binding energy to decrease. 
	
	This reveals that for a positive trion to remain bound, the system requires a tightly bound initial exciton in real space along with a spread out spatial distribution for the second doping-induced hole which shields the hole-hole interaction. As the flat-band regime is approached, these requirements are naturally met:  the reduction in the average kinetic energy causes $a_1$ to decrease, which in turn effectively screens the charge of the first positive hole.  Therefore, the second hole is shielded from the direct hole-hole repulsion, resulting in a bound positive trion.
	
\begin{figure}
		\centering
		\includegraphics[width=0.75\linewidth]{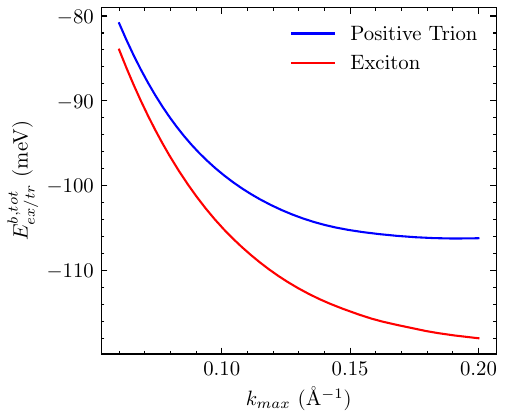}
		\caption{Plot of exciton and trion total binding energies, $E^{b,tot}_{ex/tr}$, for changing Mexican-hat width $k_{max}$ for a fixed $\Delta$ of $0.021$ eV.}
		\label{fig:changingwidth-2}\end{figure}

	We now consider the effect of varying the Mexican-hat width, $k_{max}$, with fixed $\Delta =0.021$ eV on the trion binding energy, see Fig. \ref{fig:changingwidth-2}. As $k_{max}$ increases, the binding energy of the quasiparticles also increases.  This indicates that the smaller the ring singularity of the Mexican-hat system, the larger the trion binding energy becomes for a given Mexican-hat depth. This would be maximal at the transition to the HOVHS when the ring singularity collapses to the single momentum-space point.
	Importantly, here the total (overall) binding energies of the two- and three-particles binding energies are increasing, but most notably they are increasing at a different rate. Which subsequently implies that the larger the width the larger the total binding energies become, but the lower the relative trion binding energy is. One would expect this due to the number of available states close to the band edge increasing with the increase in the Mexican-hat width, leading to an increase in binding energies.
	
	Thus in the class of type  III–VI and  IV–V 2D semiconductors, where the valence band dispersion can be described approximately by the Mexican hat dispersion, Eq. \ref{eq:mexhatdisp}, as the quadratic term, $A$, reduces to zero, the band edge and the point at which the determinant of the Hessian vanishes move closer together in $k$-space and approach $\Gamma$-point. Therefore, we find that as $A$ reduces, the positive trion becomes more favourable and it is most favourable when there is a HOVHS at the $\Gamma$-point when $A\approx 0$ as for $\alpha$-SnAs.
	
	\section{Discussion}
	This work reveals the strong influence of conventional and high-order Van Hove singularities in the valence band of 2D semiconductors on the properties of many-body excitonic and trionic states.  
	
	We first demonstrated that point-like singularities (conventional logarithmic saddles and higher-order monkey saddles) are directly mirrored in the energy dispersions of the bound states.  We found both analytically and numerically that the characteristic power-law divergence of the singularities are preserved in these cases along with the associated enhancement of the DOS for the HOVHS.  The creation and imprinting of this type of singularity in the quasi-particle DOS could open new ways to enhance the optoelectronic and light-matter interactions in these materials.  In systems with inverted Mexican hat dispersions, which host 1D line singularities, we found a critical requirement for the exciton spatial radius so that the 1D line singularity is mirrored in the exciton dispersion. Specifically, the line singularity is lost if the exciton is too tightly bound in real space.  
	
	We considered monolayer GaSe, InSe and $\alpha$-SnAs as benchmark materials, and showed how the line-singularity in this class of materials impacts the energy landscape and DOS of the trion and exciton states.  We find the most favourable positive trion state occurs when there is a HOVHS at the $\Gamma$-point due to an interplay between the interplay and balance between the electrostatic interactions and the kinetic energy of the quasi-particles. As a result, we find a positive trion state in $\alpha$-SnAs, which is bound relative to the exciton (larger total binding energy), unlike what is seen in GaSe and InSe.  

	These findings introduce new directions to engineer these exciton and trion dispersion relations, with properties that enable new optoelectronic and quantum mechanical functionalities.

	\section{Methods}
	
	\subsection{Computational Details}\label{Append-1}
	We employ the variational technique to find the ground-state quasi-particle dispersion.  We minimise the energies of the exciton and trion (singlet) to obtain an upper bound on their ground state binding energy \cite{burke2025brightening}. 
	The energy eigenvalue equation of the exciton is:\begin{equation}\begin{split}
			E_{ex} \phi_{\alpha,\beta}({\bm k}) &= [E_c({\bm k}) - E_v({\bm k} - {\bm Q})] \phi_{\alpha,\beta}({\bm k}) \\
			&+\frac{1}{A} \sum_{{\bm q}}\phi_{\alpha,\beta}({\bm k}-{\bm q})V({\bm q})
			\label{eq-BSE2}
	\end{split}\end{equation}
	where $E_{c/v}(\bm{k})$ are the conduction/valence band energies and $V(\bm{q})$ is the Rytova-Keldsh Coulomb interaction \cite{rytova2018screened,Keldysh}:\begin{equation}V({\bm q})=\frac{-2\pi e^2}{\sqrt{\kappa_z \kappa_{||}}}\frac{1}{|{\bm q}|(1+r_0|{\bm q}|)},
		\label{Eq-RKpotential}
	\end{equation}
	and $r_0$ is the screening length:
	$r_0=d(\sqrt{\varepsilon_z \varepsilon_{||}}-1)/({2\sqrt{\kappa_z \kappa_{||}}})$,
	We use the values of monolayer InSe described in Ref. \cite{Falko2020} in this work.
	We assume a hydrogenic form of the variational wavefunction so that for the ground state exciton we use the 1s hydrogenic wave function, which in momentum space is:\begin{equation}\phi_{\alpha,\beta}({\bm k})=\frac{1}{(\alpha^2+k_x^2 +k_y^2/\beta^2 )^{3/2}}
		\label{eq-1s_anisotropic}
	\end{equation}
	where $\alpha$ is the unknown variational parameter. For the exciton, $\alpha$ is the inverse exciton radius and $\beta$ is an anisotropy parameter. 

	For the positive trion, in the regime of low-doping levels, the few-body expression for the energy of the positive trion, $E_{tr,+}({\bm Q})$, reads as  \cite{burke2025brightening}: 
	\begin{equation}
		\begin{split}
			E_{tr,+}({\bm Q})\phi_{tr}({\bm k_1},{\bm k_2})= \\
			\bigg[ E_{c}({\bm k_1} +{\bm k_2})
			-E_{v_1}({\bm k_1}-{\bm Q}) - E_{v_2}({\bm k_2})\bigg] \phi_{tr}({\bm k_1},{\bm k_2})  \\
			+\sum_{{\bm q}}V({\bm q})\phi_{tr}({\bm k_1}-{\bm q},{\bm k_2}) 
			+\sum_{{\bm q}}V({\bm q})\phi_{tr}({\bm k_1},{\bm k_2}-{\bm q}) \\
			-\sum_{{\bm q}}V({\bm q})\phi_{tr}({\bm k_1}+{\bm q},{\bm k_2}-{\bm q})
		\end{split}
	\end{equation}
	where the trial wavefunction $\phi_{tr}({\bm k_1},{\bm k_2})$ due to exchange symmetry \cite{courtade2017charged} can be taken as a linear combination of the product of the two electron-hole pairs formed in the system:
	
	\begin{equation}
		\begin{split}
			\phi_{tr}({\bm k_1}&,{\bm k_2})=A(\phi_{\alpha,\beta}({\bm k_1})\phi_{\gamma,\kappa}({\bm k_2}) \\
			&(-1)^S \phi_{\alpha,\beta}({\bm k_2})\phi_{\gamma,\kappa}({\bm k_1}) )    
		\end{split}
		\label{Eq-trionwavefn}
	\end{equation}
	and the factor $(-1)^S$ changes, depending on whether the system is a singlet or a triplet. Here, we focus on the singlet trion candidate of the ground state with $S=0$. As we are looking at the ground state trion, each $\phi(\bm{k})$, is the previous excitonic 1s wave function considered above.
	
	The summations in this work are taken over the Brillouin zone and were computed using a Monkhorst-Pack grid \cite{monk1, monkhex}. 
	In addition, we use 
	a hexagonal structure with unit vectors ${\bm a_1}=\left(a/2,\sqrt{3}a/2\right)$ and ${\bm a_2}=\left(a/2,-\sqrt{3}a/2\right)$, with $a=3.95\textup{~\AA}$ as the lattice constant. The reciprocal primitive lattice vector vectors are  ${\bm b_1}=\left(2\pi/a,2\pi/\sqrt{3}a\right)$ and ${\bm b_2}=\left(2\pi/a,-2\pi/\sqrt{3}a\right)$.
	
	These states are calculated under the assumption that the hole doping is sufficiently large to remove an electron from the valence band maximum (VBM), thereby creating a hole. This positively charged hole can bind to the charge carriers present in the exciton, which is formed via photon absorption in the material. The corresponding energy dispersion relations for these quasiparticles are shown in the main text.

	\subsection{Calculation of Density of states (DOS)}\label{DOS_Append}
	To compute the DOS of the electronic and quasiparticle dispersions, we use the nature of the DOS defined as delta functions:\begin{equation}
		g(E)=\sum_{\bm{k}/\bm{Q}} \delta(E-E(\bm{k}/\bm{Q}))
	\end{equation}
	where $\bm{k}/\bm{Q}$ denotes whether we are dealing with the electronic band structure in wave-vector $\bm{k}$, or the quasiparticle in $\bm{Q}$. To compute this form, we broaden these delta functions via a Gaussian broadening for a broadening term denoted by $\sigma$. The DOS then reads as:\begin{equation}
		g(E)= \frac{1}{\sqrt{2\pi}\sigma}\sum_{\bm{k}/\bm{Q}} \text{exp}\left({\frac{-(E-E(\bm{k}/\bm{Q}))^2}{2\sigma^2}}\right)
	\end{equation}
	In this work, we use $\sigma=0.2$ meV as the appropriate broadening parameter for the energy ranges employed.

	\section{Acknowledgements}
	We would like to thank Bob Joynt, David Perkins and Ioannis Rousochatzakis for useful discussions.  The work has been supported by the EPSRC grants EP/T034351/1 and EP/X012557/1.
	\section{Author Contributions}
	LJB performed the calculations, MTG and JJB initiated and supervised the project. All authors discussed the results and wrote the paper.
	\section{Competing Interests}
	The authors declare no competing interests.
	\section{Data Availability}
	Data sets generated during the current study are available from the corresponding author on reasonable request.
	\section{Code Availability}
	The underlying code for this study is not publicly available, but may be made available to qualified researchers on reasonable request from the corresponding author.
	\bibliography{sn-bibliography.bib}

\end{document}